\documentclass[aps,amsmath,amssymb,twocolumn,longbibliography,10pt,pra]{revtex4-1}
\usepackage[T1]{fontenc}
\usepackage{graphicx}
\usepackage[colorlinks=true,linkcolor=blue,citecolor=red,urlcolor=magenta]{hyperref}
\usepackage{braket}

\begin{document} 
\title{Unconventional pairing in one-dimensional systems \\ of a few mass-imbalanced ultracold fermions}
\author{Patrycja {\L}yd{\.z}ba}
\email{patrycja.lydzba@pwr.edu.pl}
\affiliation{Department of Theoretical Physics, Faculty of Fundamental Problems of Technology, Wroclaw University of Science and Technology, Wyb. Wyspia\'{n}skiego 27, 50-370, Wroc{\l}aw, Poland}

\author{Tomasz Sowi\'{n}ski}
\email{tomsow@ifpan.edu.pl}
\affiliation{Institute of Physics, Polish Academy of Sciences, Aleja Lotnik\'{o}w 32/46, 02-668, Warsaw, Poland}
\date{\today} 

\begin{abstract}
We study the ground-state properties of a two-component fermionic mixture effectively confined in a one-dimensional harmonic trap. We consider scenarios when numbers of particles in components are the same but particles have different masses. We examine whether it is possible to detect signatures of an unconventional pairing between opposite-spin fermions in the presence of attractive interactions. For this purpose, we perform the exact diagonalization of the many-body Hamiltonian and study the two-particle reduced density matrix. In agreement with expectations, we confirm that the many-body ground state is dominated by conventional pairs with a negligible total momentum for a small mass imbalance. Furthermore, we show that for sufficiently large mass ratios the domination of fundamentally different pairs is established and the Fulde-Ferrell-Larkin-Ovchinnikov phase is supported. Finally, we argue that the two mechanisms can coexist in the regime of moderate mass ratios. Due to the current experimental progress in obtaining ultra-cold fermionic systems in a few-body regime, our predictions may have some importance for the upcoming experiments.
\end{abstract}

\maketitle
 
\section{Introduction}
The discovery of superconductivity by Onnes \cite{1911Onnes} provided one of the most challenging phenomena for a theoretical explanation. After many unsuccessful attempts and failed theoretical propositions \cite{2010CooperFeldmanBook}, Cooper made a luminous observation that the effective attraction between opposite-spin particles can lead to the spontaneous formation of correlated two-particle states above the Fermi sea \cite{1956CooperPhysRev}. This opened a path towards the appropriate theory of superconductivity, which was finally formulated by Bardeen, Cooper, and Schrieffer \cite{1975BardeenPhysRev}. In the simplest case of a balanced mixture of fermions, paired particles have opposite spins and momenta (\textit{i.e.,} a negligible center-of-mass momentum). The pairing is possible since the Fermi spheres of both components are identical. More specifically, single-particle excitations from one surface can be adjusted to excitations from another surface. Although the initial theory concerned electrons moving in metallic crystals, it quickly became apparent that the pairing mechanism is fundamental and present in various strongly-correlated systems, like heavy nuclei \cite{1958BohrPhysRev,1959MigdalNucPhys}, neutron stars \cite{1969BaymScience,1971YangNucPhysA} (for comprehensive review see \cite{2019SedrakianEPJA}), or ultra-cold mixtures of fermions \cite{2004ViveritPRL,2004JullietPRL,2008GiorginiRevModPhys}. What is more, footprints of correlated pairs have been experimentally found in the ultra-cold system of a few $^6$Li atoms confined in the almost perfect one-dimensional harmonic trap \cite{2013ZurnPRL}. These experimental findings are consistent with the theoretical description in terms of the Cooper pairing mechanism \cite{2015SowinskiEPL,2015DamicoPRA,2016HofmannPRA}.

The picture is more complicated when components do not comprise the same numbers of particles and the Fermi surfaces are not compatible. Then, as independently predicted by Fulde and Ferrell \cite{1964FuldePhysRev} and Larkin and Ovchinnikov \cite{1965LarkinJETP} (FFLO), the formation of correlated pairs requires a relative shift of the Fermi spheres. Consequently, the pairing with a non-zero total momentum is supported. It is currently believed that both phases (\textit{i.e.}, the standard Cooper pairing and the FFLO pairing) can be established in the same regime of parameters, but the FFLO phase is energetically favorable \cite{2018KinnunenRPP}. Although the latter still awaits a clear experimental evidence, recent results of measurements in ultra-cold quantum gasses are promising \cite{2010LiaoNature}. Moreover, it was shown theoretically that the FFLO phase is supported and should be detectable in one-dimensional spin-imbalanced fermionic systems containing only a few atoms \cite{2019Pecak2}.

It has been argued that the FFLO mechanism of pair formation is also present in two-component systems with balanced numbers of particles but different masses \cite{2003LiuPRL,2007Kaczmarczyk}. This scenario is rather rare in standard solid-state systems since it can only be achieved effectively by forcing the component-dependent dispersion relations \cite{2005Collam,1995Korbel}. On the other hand, it can be obtained directly in ultra-cold atomic systems by preparing mixtures of atoms with different masses, like lithium-potassium \cite{2008WillePRL,2010TieckePRL,2016CetinaScience} or dysprosium-potassium \cite{2018RavensbergenPRA}. Although the transitions to unconventional superconducting phases in systems with a non-zero polarization have been thoroughly investigated using different methods \cite{2007Orso,2011Feiguin,2017Ptok,2017Franca,2018Franca}, they are less explored in systems with a non-zero mass imbalance \cite{2010Orso}.

All the above facts motivated us to further explore the properties of attractively-interacting fermionic mixtures with a mass imbalance. In this paper, we focus on a particular and exotic system, in which specific signatures of the FFLO phase can be captured. Namely, we focus on a few ultra-cold fermions confined in a one-dimensional external trap. We analyze their properties mostly in terms of the Penrose-Onsager criterion for the condensation of correlated pairs \cite{1956PenrosePhysRev,2004JullietPRL}. More specifically, we study the occupations and momentum correlations of dominant orbitals of the two-particle reduced density matrix. In this way, we extend recent results on the Cooper-like pairing in mass-balanced systems \cite{2015SowinskiEPL} and in mass-imbalanced systems \cite{2019PecakPRA}. In the latter case, the shot-noise correlations and inter-component entropies were examined.

\section{The system}
In our theoretical model, we consider a two-component system of interacting fermions moving in a one-dimensional harmonic trap of frequency $\omega$. We assume that the trap is identical for both components containing fundamentally different atoms (with a different mass, spin, {\it etc.}). This is encoded in a generalized quantum number $\sigma\in\left\{\uparrow,\downarrow\right\}$. The many-body Hamiltonian can be written in the real-space representation as
\begin{align}
\label{eq1}
\hat{H} &= \sum_\sigma\!\int\!\!\mathrm{d}x\,\hat{\Psi}^\dagger_{\sigma}\left(x\right) h_\sigma \hat{\Psi}_{\sigma}\left(x\right)+ \nonumber \\
&+ g\int\!\!\mathrm{d}x\,\hat{\Psi}^\dagger_{\uparrow}(x) \hat{\Psi}^\dagger_{\downarrow}(x) \hat{\Psi}_{\downarrow}(x) \hat{\Psi}_{\uparrow}(x)
\end{align}
where $h_\sigma$ is the single-particle Hamiltonian of the component $\sigma$ comprising kinetic and potential energy terms
\begin{equation}
h_\sigma = -\frac{\hbar^2}{2m_\sigma}\frac{\mathrm{d}^2}{\mathrm{d}x^2}+\frac{m_\sigma\omega^2}{2}x^2.
\end{equation}
We assume that interactions between particles from different components can be modeled with a zero-range $\delta$-like potential of strength $g$. This approximation is reasonable and relevant to experiments in ultra-cold quantum gases \cite{2019SowinskiRPP}. It should be emphasized that the methods originating in the Feshbach resonance phenomenon and based on varying the external confinement in perpendicular directions allow to tune the effective one-dimensional scattering length between atoms, and consequently the strength of contact interactions in these systems \cite{PethickBook,1998OlshaniiPRL,2012ZwergerBook}. These interactions are prohibited for particles with the same spin by the Pauli exclusion principle. Let us mention that $\hat{\Psi}_{\sigma}\left(x\right)$ is a fermionic field operator, which annihilates a $\sigma$-spin fermion at position $x$ and obeys conventional fermionic anti-commutation relations,
\begin{equation}
\begin{split}
& {\left\{\hat{\Psi}_{\sigma}(x),\hat{\Psi}_{\sigma'}(x')\right\}=0},\\
& {\left\{\hat{\Psi}^{\dagger}_{\sigma}(x),\hat{\Psi}_{\sigma'}(x')\right\}=\delta_{\sigma\sigma'}\delta(x-x')}.
\end{split}
\end{equation}
Furthermore, the many-body Hamiltonian \eqref{eq1} commutes with the number operators, $\hat{N}_{\sigma} = \int\!\mathrm{d}x\;\hat{\Psi}^{\dagger}_{\sigma}(x)\hat{\Psi}_{\sigma}(x)$. As a result, the two-component mixture can be studied in the independent subspaces in which populations of atoms, $N_\uparrow$ and $N_\downarrow$, are individually fixed. In the following, we consider a balanced number of particles ($N_\uparrow=N_\downarrow=N/2$) and attractive interactions ($g<0$). 

For a convenience, we rewrite all quantities in the dimensionless units, {\it \textit{i.e.}}, we express all energies in $\hbar\omega$, lengths in the harmonic oscillator length of a spin-up particle $\sqrt{\hbar/(m_{\uparrow}\omega})$ and the interaction strength in $\sqrt{\hbar^3\omega/m_{\uparrow}}$. With this convention, the mass $m_\uparrow$ is set to unity and only the mass ratio $\mu=m_\downarrow/m_\uparrow$ is a relevant parameter. Without losing generality, we assume that $\mu\geq 1$.

In this work we investigate the ground-state properties of a two-component fermionic system in the exact diagonalization approach. First, we rewrite the many-body Hamiltonian \eqref{eq1} in a convenient representation, in which a single-particle basis comprises eigenstates of $h_\sigma$,
\begin{equation} \label{HamRep2}
\hat{H} = \sum_{i\sigma} \epsilon_i \hat{a}^{\dagger}_{i\sigma} \hat{a}_{i\sigma} + g\sum_{i,j,k,l} I_{ijkl} \hat{a}^{\dagger}_{i\uparrow} \hat{a}^{\dagger}_{j\downarrow} \hat{a}_{k\downarrow} \hat{a}_{l\uparrow},
\end{equation}
where $\epsilon_i = \left(i+\frac{1}{2}\right)$ is a single-particle energy independent of spin and $\hat{a}_{i\sigma}$ is the operator annihilating a $\sigma$-spin particle from the $i$-th harmonic oscillator state described by the wave function $\phi_{i\sigma}(x)$. Furthermore, 
\begin{equation}
I_{ijkl} = \int \mathrm{d}x\,\phi^*_{i\uparrow}(x)\phi^*_{j\downarrow}(x)\phi_{k\downarrow}(x)\phi_{l\uparrow}(x). 
\end{equation}
The representation \eqref{HamRep2} is obtained, when a decomposed field operator,
\begin{equation}
\hat{\Psi}_{\sigma}\left(x\right) = \sum_i \phi_{i\sigma}(x) \hat{a}_{i\sigma},
\end{equation}
is introduced into \eqref{eq1}. Next, we perform the exact diagonalization of the many-body Hamiltonian using the implicitly restarted Arnoldi method \cite{1998ARPACK}. For this purpose, we restrict the single-particle basis to the $\cal K$ lowest-energy eigenstates of $h_\sigma$. As discussed in \cite{2019PecakPRA}, ${\cal K}=10$ provides reliable results for one-dimensional systems with at most $N_\sigma=5$ particles and interactions considered here. We independently confirm this observation in Fig.~\ref{fig1} and Fig.~\ref{fig2}, where we reproduce the most important results for different cut-offs $\cal K$ and determine how they are affected by the increased $\cal K$. Finally, we obtain the lowest-energy eigenstate $|\mathtt{G}_0\rangle$ of the many-body Hamiltonian $\hat{H}$ and exploit it to determine the ground-state properties of a two-component fermion system.

\section{Varying the mass ratio}
We aim to determine the ground-state properties of a two-component fermion system for different mass ratios. In particular, we want to establish whether the conventional paired phase undergoes a transition to any of unconventional paired phases ({\it \textit{e.g.}}, the fragmented phase with more than one dominant eigenvalue of a two-body reduced density matrix \cite{2016SakmannNatPhys} or the FFLO phase with Cooper-like pairs moving with a non-zero center-of-mass momentum \cite{2010TezukaNJP,2018PtokJSupNovMag}). It is not trivial to describe the condensation of Cooper-like pairs in a one-dimensional system of a few fermions, since the off-diagonal long-range order is replaced by the power-decaying law \cite{2010TezukaNJP,2018PtokJSupNovMag}. As a result, the mean-field considerations are of a limited benefit. Nevertheless, some research in this direction has been performed. For example, the conventional paired phase has been studied in terms of a two-particle reduced density matrix of opposite-spin fermions in the mass balanced scenario \cite{2015SowinskiEPL}. In the case of atoms with a different mass, it has been demonstrated that the system displays an apparent spatial separation of components for repulsive interactions \cite{2016PECAK}. Simultaneously, the inter-component correlations are quickly suppressed with increasing $\mu$ for attractive interactions. All it means that the unconventional paired phase can be dominant in a narrow, if any, regime of mass ratios $\mu$. In the following, we examine this possibility.

The pairing of opposite-spin fermions, as a two-particle correlation phenomenon, is encoded in a two-particle reduced density matrix,
\begin{equation} \label{2rdm}
\rho_2(ij;kl) = \frac{1}{N_\uparrow N_\downarrow} \bra{\mathtt{G}_0} \hat{a}^{\dagger}_{i\uparrow}\hat{a}^{\dagger}_{j\downarrow} \hat{a}_{l\downarrow}\hat{a}_{k\uparrow} \ket{\mathtt{G}_0}.
\end{equation}
This density matrix can be expressed in a diagonal form, {\it \textit{i.e.}}, as a linear combination of projection operators onto its eigenstates $|\lambda_\alpha\rangle$,
\begin{equation} \label{decomposition}
{\rho}_2=\sum_\alpha \lambda_\alpha \ket{\lambda_\alpha}\bra{\lambda_\alpha}.
\end{equation}
The probabilities that eigenstates are occupied correspond to $\lambda_\alpha$ and satisfy the condition $\sum_\alpha \lambda_\alpha=1$. Two-particle orbitals $\ket{\lambda_\alpha}$ can be represented by corresponding wave functions $\Lambda_\alpha(x,y)$ and $\Pi_\alpha(p,k)$ in the position and the momentum representation, respectively. They are defined as
\begin{subequations}
\begin{align}
\Lambda_\alpha(x,y) = \sum_{ij}C^{(\alpha)}_{ij} \phi_{i\uparrow}(x)\phi_{j\downarrow}(y), \\
\Pi_\alpha(p,k) = \sum_{ij}C^{(\alpha)}_{ij} \widetilde{\phi}_{i\uparrow}(p)\widetilde{\phi}_{j\downarrow}(k).
\end{align}
\end{subequations}
Here $C^{(\alpha)}_{ij}$ are the decomposition coefficients of a two-particle orbital $\ket{\lambda_\alpha}=\sum_{ij}C^{\alpha}_{ij}\,\hat{a}_{i\uparrow}^\dagger\hat{a}_{j\downarrow}^\dagger|\mathtt{vac}\rangle$ obtained from the diagonalization of the density matrix \eqref{2rdm}, while $\widetilde{\phi}_{i\sigma}(p)=\int\mathrm{d}x\phi_{i\sigma}(x)\mathrm{exp}(-ipx)$ is the Fourier transform of a wave function of a single-particle orbital.

%Figure 1______________________________
\begin{figure}[h!]
\centering
\includegraphics[width=\linewidth]{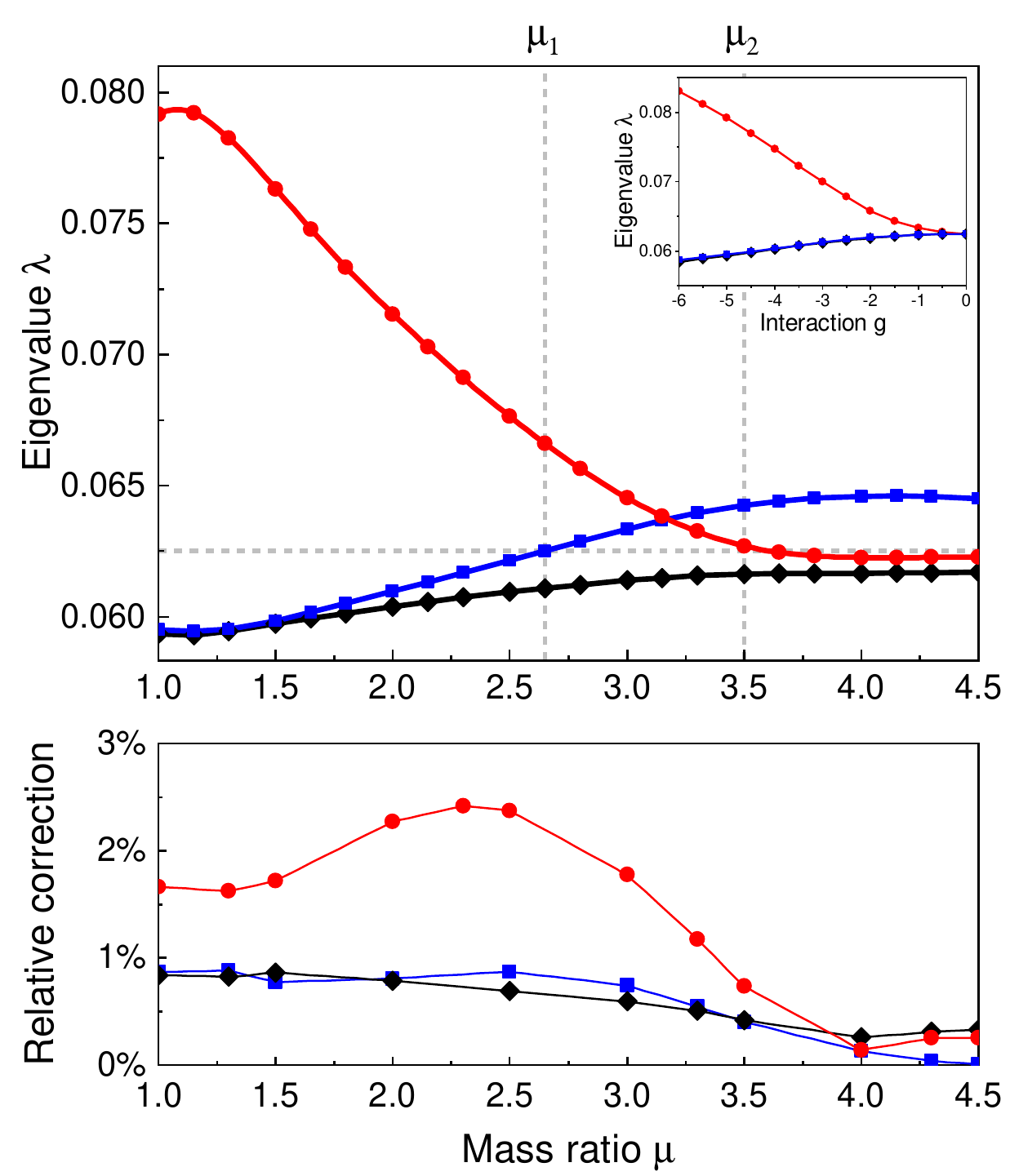}
\caption{\label{fig1} {\bf (top)} Spectrum of a two-body reduced density matrix for a system with $N=8$ particles. Red (points), blue (squares) and black (diamonds) lines correspond to the three largest eigenvalues $\lambda_0$, $\lambda_1$ and $\lambda_2$, respectively. The horizontal line marks the non-interacting value $(N_\uparrow N_\downarrow)^{-1}$. The first vertical line marks the mass ratio $\mu_1$ for which the approximation of $\rho_2$ by a single projection operator $|\lambda_0\rangle\langle\lambda_0|$ is no longer valid, while the second line marks the mass ratio $\mu_2$ for which the dominance of an eigenvalue is recovered but in another orbital $|\lambda_1\rangle$. We suspect that a different pairing mechanism is supported in this regime of parameters (\textit{i.e.}, for $g\approx -5$ and above $\mu_2$). Note that it is less effective than the conventional mechanism (\textit{i.e.}, fewer particles form Cooper-like pairs) and it is related to a different eigenstate of a two-body reduced density matrix $|\lambda_1\rangle$. {\bf (inset)}  Eigenvalues $\lambda_\alpha$ of the two-body reduced density matrix as functions of interactions $g$ for a mass-balanced situation $\mu=1$. {\bf (bottom)} The relative corrections $|\lambda_{{\cal K}+1}-\lambda_{\cal K}| / \lambda_{\cal K}$ for the three largest eigenvalues introduced after increasing the cut-off from ${\cal K}=10$ to ${\cal K}=11$ (colors correspond to upper panel). Note that their values are below $2.5\%$ for the entire range of mass ratios. It means that the spectrum of the density matrix $\rho_2$ is almost unaffected by the increased cut-off ${\cal K}$ and it can still be divided into the three regimes separated by the vertical grey lines. The lines are only slightly shifted when a greater ${\cal K}$ is used. This observation confirms the validity of our results. Interaction strength is expressed in natural units of the harmonic oscillator $\sqrt{\hbar^3\omega/m_{\uparrow}}$.}
\end{figure}
%Figure 1______________________________

Along with the Penrose-Onsager criterion \cite{2003Shi,2003Shi2}, if a certain eigenvalue $\lambda_0$ dominates in the decomposition \eqref{decomposition}, the two-particle reduced density matrix ${\rho}_2$ can be approximated by a single projection operator $\ket{\lambda_0}\bra{\lambda_0}$. Namely, the system can be described as a condensate of pairs occupying orbital $\ket{\lambda_0}$. As a result, the collective pairing of opposite-spin fermions is characterized by a significant domination of a certain orbital over other orbitals. 

To make further analysis as clear as possible, we first investigate eigenvalues of a two-body reduced density matrix $\rho_2$ for a mass-balanced system ($\mu=1$). When interactions are negligible (\textit{i.e.}, in the limit $g\rightarrow 0$), exactly $N_\uparrow N_\downarrow$ eigenvalues equal to $(N_\uparrow N_\downarrow)^{-1}$ are non-zero. However, only one increases with increasing attraction $g$ and quickly dominates the spectrum (\textit{i.e.}, $\lambda_0$). This behavior of the eigenvalues has been observed previously \cite{2015SowinskiEPL} and it is in line with expectations, since atoms are more willing to form pairs when the attraction is stronger (see inset in Fig.~\ref{fig1}).

Subsequently, we select the interaction strength for which the dominance of $\ket{\lambda_0}$ is significant (we take $g=-5$), and we explore the evolution of all eigenvalues in a varying mass ratio $\mu$. The latter is presented in Fig.~\ref{fig1} for the three largest $\lambda_\alpha$ and systems of $N=8$ particles. For clarity, we keep the order of $\lambda$'s as established for $\mu=1$ in Fig.~\ref{fig1} and throughout the paper. For example, regardless of their order and relative values, red and blue lines always correspond to $\lambda_0$ and $\lambda_1$, respectively. It is clearly visible that the gap $\Delta\lambda=\lambda_0-\lambda_1$ is diminishing with an increasing mass ratio $\mu$. Consequently, the approximation of a two-body reduced density matrix by a single projection operator becomes less justified for larger mass ratios $\mu$ until it completely breaks down. Eventually, when $\mu>\mu_1\approx 2.5$ is exceeded (marked by the first vertical grey line in Fig.~\ref{fig1}), the second largest eigenvalue surpasses its non-interacting value $(N_\uparrow N_\downarrow)^{-1}$ (marked by the horizontal grey line in Fig.~\ref{fig1}). Therefore, the system can no longer be described as a condensate of pairs in $\ket{\lambda_0}$. A similar transition was observed in previous studies in terms of inter-component correlations \cite{2019PecakPRA}. The situation is slightly different when the mass ratio $\mu$ is further increased. It turns out that above $\mu_2\approx 3.5$ (marked by the second vertical grey line in Fig.~\ref{fig1}), the initially largest eigenvalue $\lambda_0$ drops below its non-interacting value $(N_\uparrow N_\downarrow)^{-1}$ and the dominance of the initially second eigenstate $|\lambda_1\rangle$ is established. It is worth noting that in this regime the dominance is much weaker than in the regime of a small mass-imbalance. Furthermore, the order of eigenvalues is changed, {\it \textit{i.e.}}, the role of a dominant orbital is taken over by another eigenstate of a two-body reduced density matrix. The occupation of a different two-particle orbital signals that another pairing mechanism can be activated. Only when this mechanism becomes inefficient, heavier atoms de-correlate from lighter atoms as argued in \cite{2019PecakPRA}.

\section{Momentum correlations}

%Figure 2______________________________
\begin{figure}
\centering
\includegraphics[width=\linewidth]{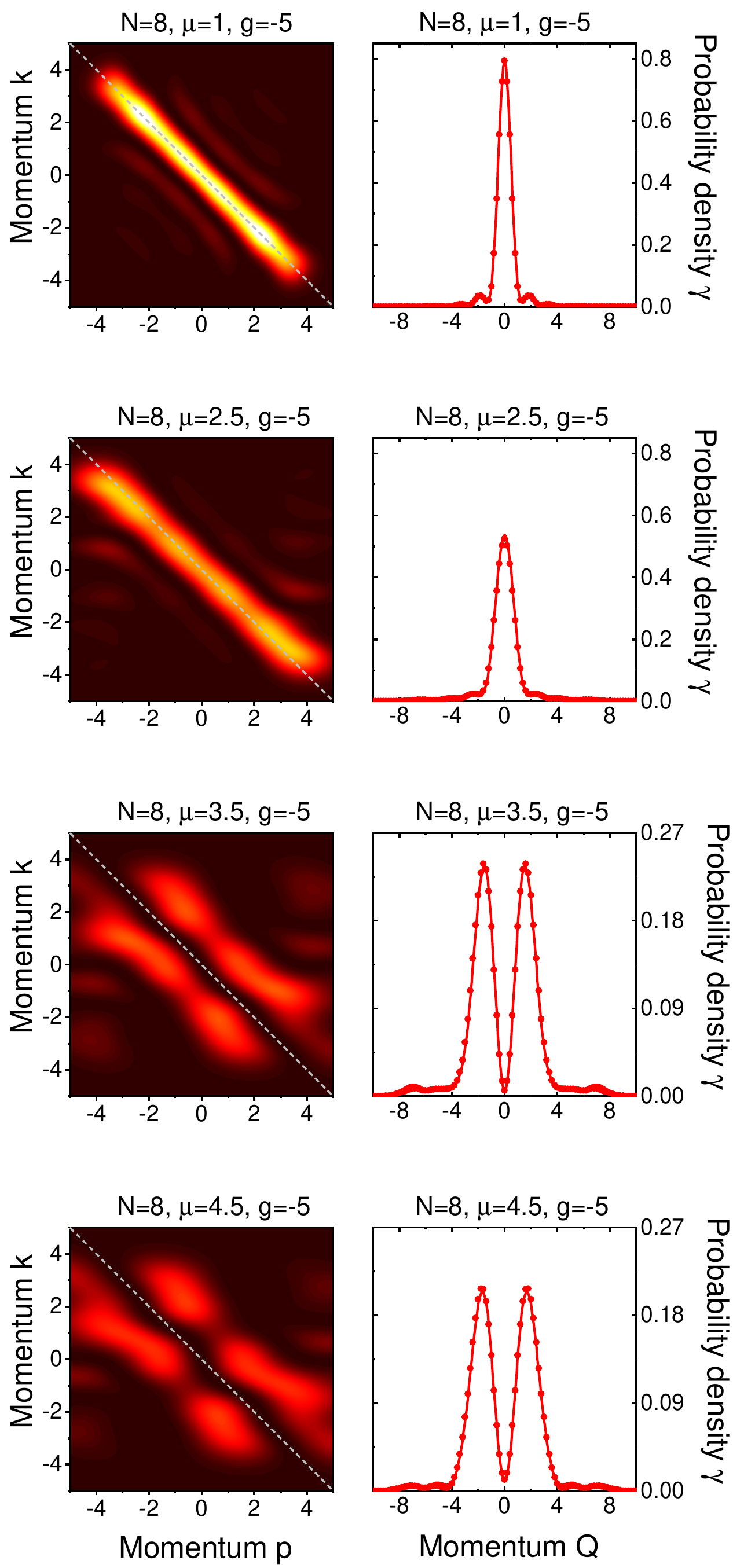}
\caption{\label{fig2} {\bf (left column)} The probability density of finding a pair of atoms occupying the dominant orbital $\ket{\lambda_\alpha}$ in a momentum configuration $|\Pi_\alpha(p,k)|^2$. For a minor mass imbalance, the distribution is non-vanishing only in a close vicinity of a diagonal line $k=-p$ (marked by a gray dashed line). Simultaneously, for a major mass imbalance, the distribution develops two elongated maxima. They are located on both sides and equally spaced from a diagonal line $k=-p$. {\bf (right column)} The probability density integrated over a momentum of a single atom. Red solid lines and red dots correspond to results obtained for cut-offs ${\cal K}=10$ and ${\cal K}=11$, respectively. Note their precise agreement proving a satisfactory convergence of the results. See the main text for details. All momenta and probability densities are measured in natural units of the harmonic oscillator, $\sqrt{\hbar m_\uparrow \omega}$ and $1/\sqrt{\hbar m_\uparrow \omega}$, respectively.}
\end{figure}
%Figure 2______________________________

If the hypothesis outlined above is correct and another pairing mechanism is supported in a two-component fermion system above $\mu_2$, it should manifest in the correlations between momenta of atoms occupying the dominant orbital (when $\ket{\lambda_1}$ starts to dominate over $\ket{\lambda_0}$). These can be established by considering the probability density of finding a pair in a given momentum configuration $|\Pi_\alpha(p,k)|^2$. We present this distribution in the left column of Fig.~\ref{fig2} for systems comprising $N=8$ particles and different mass ratios $\mu$. Note, that due to the space reflection symmetry of the system, two-body distributions are point-reflection symmetric, $|\Pi_\alpha(p,k)|^2=|\Pi_\alpha(-p,-k)|^2$. For a minor mass imbalance (when $\ket{\lambda_0}$ dominates) the distribution $|\Pi_0(p,k)|^2$ is non-vanishing only in a close vicinity of a diagonal line $k=-p$ (marked by the grey line in Fig.~\ref{fig2}). Furthermore, it slowly spreads over the configuration space with increasing $\mu$, and finally slightly shifts towards a non-zero center-of-mass momentum for large values of $k$ and $p$. Recall that the off-diagonal long-range order is replaced by the power decaying law ({\it \textit{i.e.}}, the algebraically decaying off-diagonal long-range order) in one-dimensional systems \cite{2008Batrouni}. Therefore, a perfect anti-correlation between momenta of opposite-spin fermions is not expected.

\begin{figure}
\includegraphics[width=\linewidth]{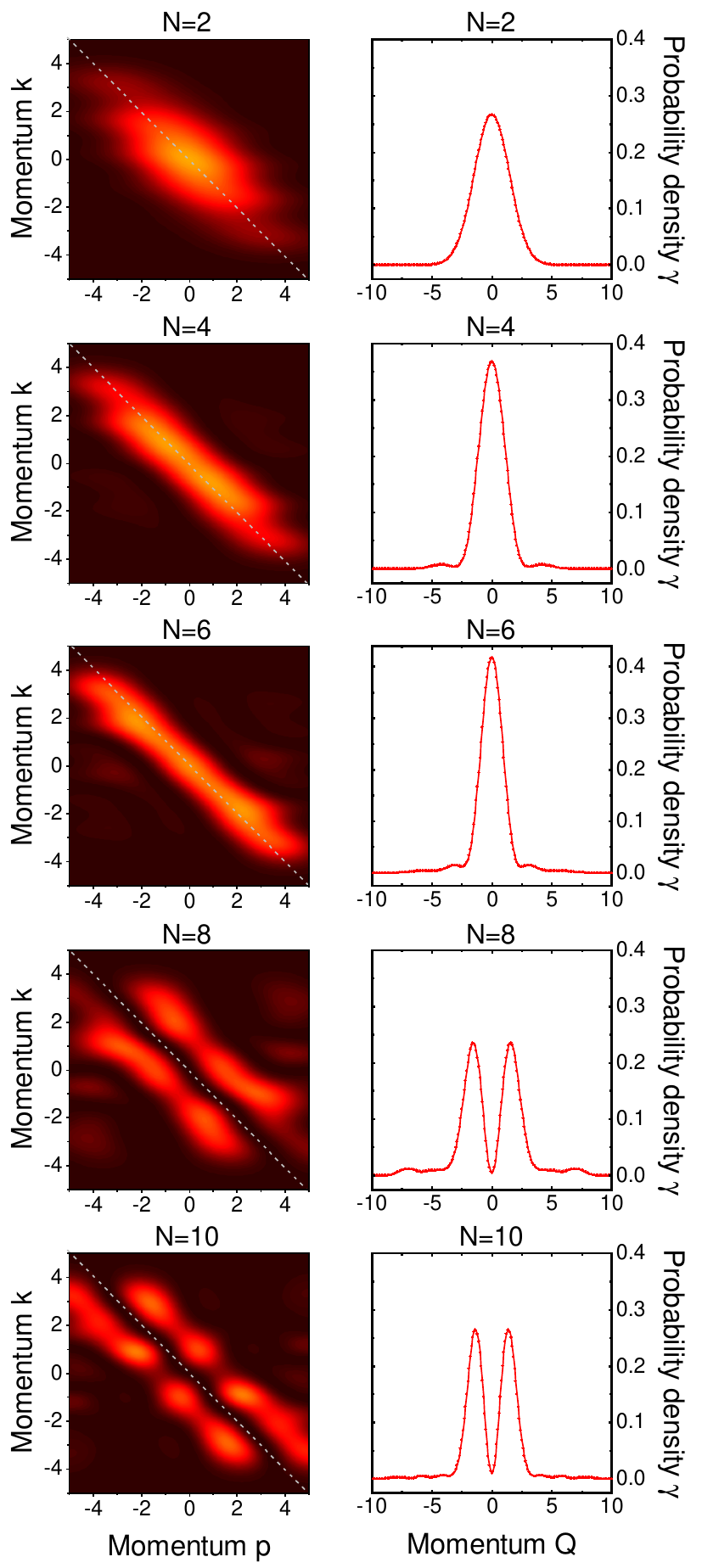}
\caption{\label{fig2b} The comparison of probability densities of the dominant orbital $|\Pi_\alpha(p,k)|^2$ for attractively interacting systems $g=-5$ with a large mass imbalance $\mu=3.5$ which comprise $N=2,\dots,10$ particles. Maxima located on both sides of a diagonal line $k=-p$ signal the appearance of the Fulde-Ferrell-Larkin-Ovchinnikov phase for more than $N=6$ particles. All momenta and probability densities are measured in natural units of the harmonic oscillator, $\sqrt{\hbar m_\uparrow \omega}$ and $1/\sqrt{\hbar m_\uparrow \omega}$, respectively. }
\end{figure}

In contrast, the probability density is abruptly modified for a major mass imbalance ($\mu>\mu_2$ where $\ket{\lambda_1}$ is dominating). Strictly speaking, it is negligible for configurations with zero center-of-mass momentum, and comprises maxima located on both sides of a diagonal line $k=-p$. This modification seems to be present for systems comprising different numbers of particles and it is more pronounced for larger $N$. For a mass ratio $\mu=3.5$ and interactions $g=-5$ considered in Fig.~\ref{fig2b}, the modification is fully developed for more than $N=6$ particles. It indicates that the FFLO-like pairing mechanism is activated in a two-component fermion system \cite{1964FuldePhysRev,2018KinnunenRPP}. More specifically, Cooper-like pairs with a non-zero center-of-mass momentum, either $+Q$ or $-Q$, are formed.   

%Figure 3______________________________
%Figure 3______________________________

At this point we remark that the probability density of a dominant orbital $|\Pi_\alpha(p,k)|^2$ reflects some properties of the so-called shot-noise correlations $\mathcal{G}(p,k)$ published in \cite{2008LuscherPRA}, which can be measured in the expansion experiments \cite{2004AltmanPRA,2010Loh,2011Kajala}. As argued in \cite{2008LuscherPRA}, the separation of the distribution $\mathcal{G}(p,k)$ for a spin-imbalanced system signals the appearance of the FFLO phase (see \cite{2019Pecak2} for details). However, signatures of Cooper-like pairs with a non-zero total momentum have not been found in the shot-noise correlations of a mass-imbalanced system \cite{2019PecakPRA}. This should not be surprising since the FFLO signal is proportional to the number of paired particles (determined by the dominant eigenvalue $\lambda_1$), which is significantly smaller in the regime of a major mass imbalance. Recall that an analogous reduction has been predicted for a superconducting order parameter, at least in quasi-one-dimensional systems with a non-zero polarization \cite{2018PtokJSupNovMag,2012Kim,2008Liu}. Nevertheless, we have closely examined the shot-noise correlations published in \cite{2019PecakPRA} (see Fig.~\ref{fig2} therein). When momenta of atoms are small, the particular pattern confirming the existence of Cooper-like pairs cannot be seen in the shot-noise picture. In other words, there are no correlations between $k$ and $p$. However, two robust maxima located in the opposite corners of the configuration space can be distinguished. They are equally spaced from a diagonal line $p=-k$ and related by the inversion symmetry. Therefore, we are convinced that subtle signatures of the Fulde-Ferrell-Larkin-Ovchinnikov pairing are present in the distribution $\mathcal{G}(p,k)$, although they have not been previously recognized. This suggests that the unconventional paired phase realized in one-dimensional mass-imbalanced systems can be observed in ultracold quantum gas experiments.

\section{The total momentum of a pair}
\begin{figure}
\centering
\includegraphics[width=\linewidth]{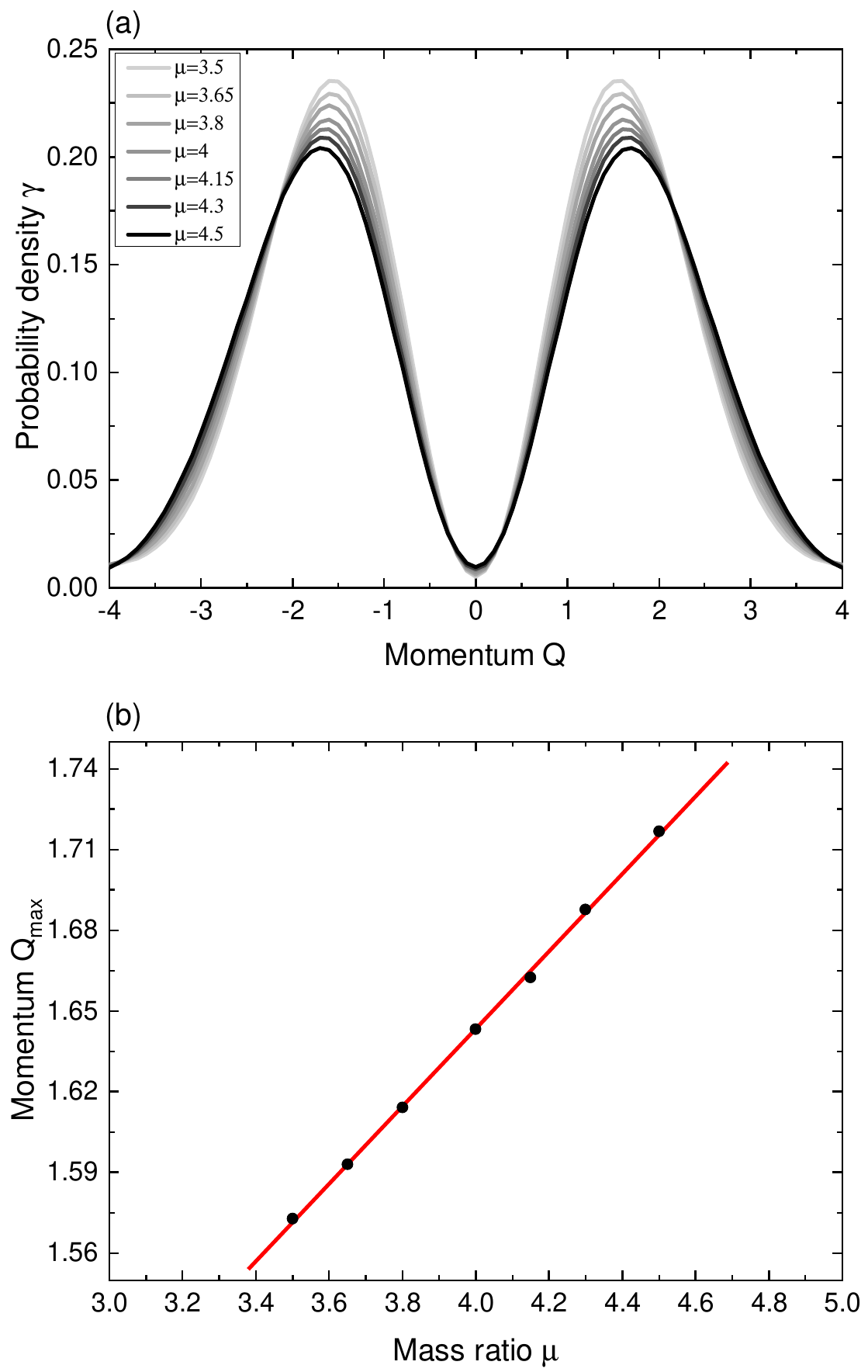}
\caption{\label{fig3} {\bf (top)} The probability density $\gamma\left(Q\right)$ for different mass ratios $\mu>\mu_2$ and $N=8$. {\bf (bottom)} The most probable FFLO momentum $Q_\mathrm{max}$ as a function of mass ratio. The linear dependence is clearly visible. All momenta and probability density are measured in natural units of the harmonic oscillator, $\sqrt{\hbar m_\uparrow \omega}$ and $1/\sqrt{\hbar m_\uparrow \omega}$, respectively.}
\end{figure}

Cooper-like pairs with a non-zero center-of-mass momentum, $Q$, are formed when opposite-spin components have incompatible Fermi surfaces \cite{2018KinnunenRPP}. In polarized systems, the incompatibility is provided by unequal particle numbers or chemical potentials of opposite-spin components. Furthermore, it has been demonstrated that the center-of-mass momentum $Q$ is proportional the difference in Fermi momenta \cite{2008Batrouni}. This can be easily translated into the dependence on particle numbers or chemical potentials. In the considered one-dimensional system placed in a harmonic trap, the definition of a Fermi momentum is not straightforward. Nevertheless, the mass imbalance leads to unequal densities of opposite-spin components \cite{2018KinnunenRPP}. As a result, the incompatibility as well as the Fulde-Ferrell-Larkin-Ovchinnikov phase, as we have demonstrated in the previous paragraph, is obtainable. It is natural to wonder whether there is a simple dependence between the center-of-mass momentum $Q$ and the mass ratio $\mu$. To find the answer to this question, we have integrated the probability density of a dominant orbital $|\Pi_\alpha(p,k)|^2$ over a momentum of a single atom. In this way, we have obtained the following function,
\begin{equation}
\gamma\left(Q\right)=\int_{-\infty}^{\infty}\!\!\mathrm{d}p\,|\Pi_\alpha(p,p-Q)|^2.
\end{equation}
The latter is presented in the right column of Fig.~\ref{fig2} for systems comprising $N=8$ particles and different mass ratios $\mu$. In line with expectations, for a minor mass imbalance $\mu<\mu_1$, the probability density $\gamma\left(Q\right)$ is strongly peaked around $Q=0$ and vanishes with increasing center-of-mass momentum. However, the peak splits into two maxima when the mass ratio exceeds $\mu_2$. These maxima are symmetrically arranged around $Q=0$. We have established that their relative distance is almost independent from the strength of interactions $g$. Simultaneously, it grows with the mass ratio $\mu$ as demonstrated in Fig.~\ref{fig3}a. Moreover, the most probable FFLO momentum $Q_\mathrm{max}$ ({\it i.e.}, the value of $Q$ for which the function $\gamma(Q)$ is maximal) fits almost perfectly to the linear function of a mass ratio $\mu$ (Fig.~\ref{fig3}b).
%Figure 4______________________________
%Figure 4______________________________
%Figure 5______________________________
\begin{figure}
\includegraphics[width=0.9\linewidth]{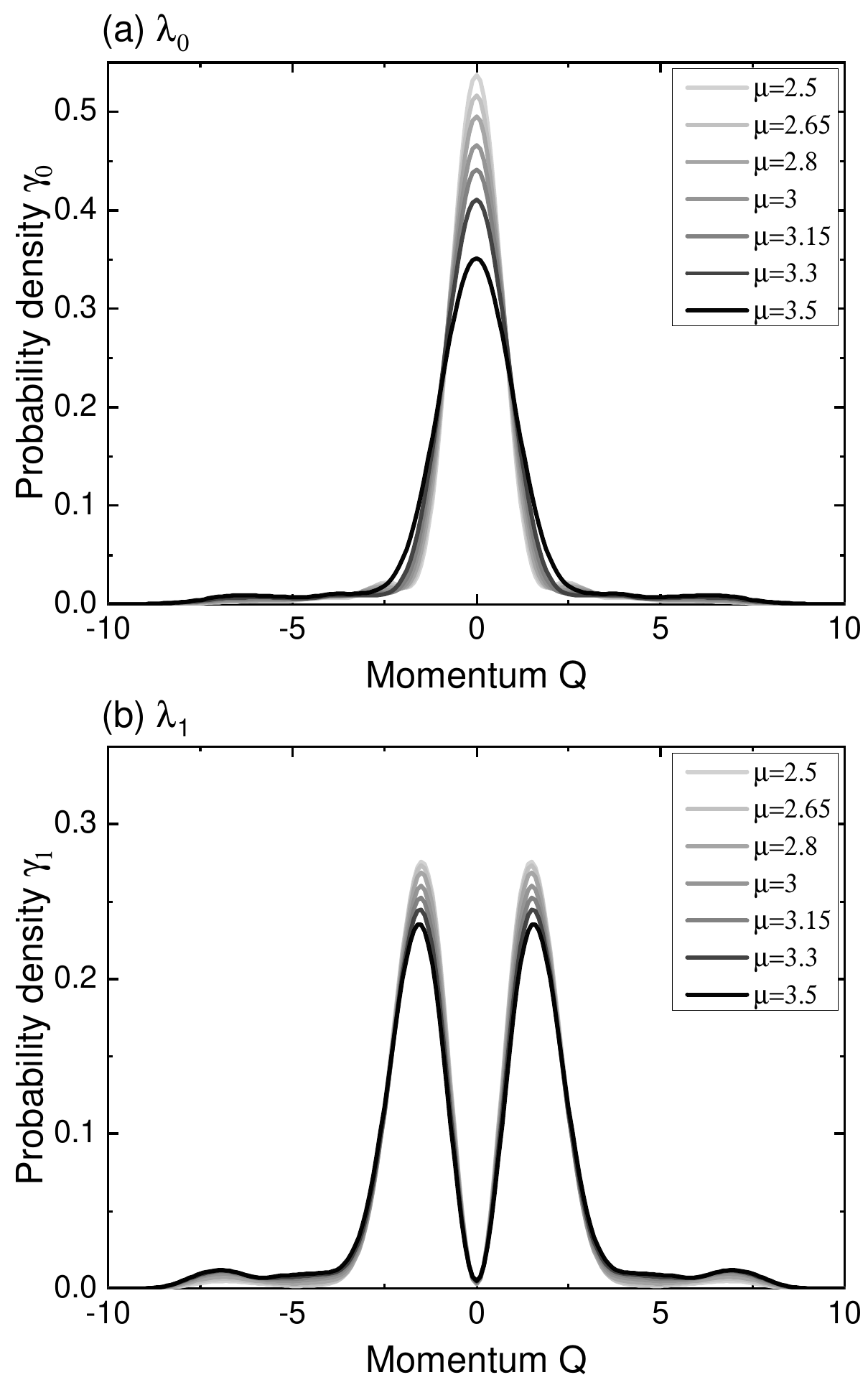}
\caption{\label{fig4} The coexistence of pairing mechanisms. The probability density $\gamma\left(Q\right)$ calculated for $\ket{\lambda_0}$ {\bf (top)} and $\ket{\lambda_1}$ {\bf (bottom)}. These two-body orbitals correspond to the two largest eigenvalues of a two-body reduced density matrix, respectively. Systems comprising $N=8$ particles for moderate mass ratios $\mu\in[2.5,3.5]$ have been considered. All momenta and probability densities are measured in natural units of the harmonic oscillator, $\sqrt{\hbar m_\uparrow \omega}$ and $1/\sqrt{\hbar m_\uparrow \omega}$, respectively.}
\end{figure}
%Figure 5______________________________

\section{The coexistence of paired phases}
According to \cite{2018PtokJSupNovMag}, the conventional paired phase and the Fulde-Ferrell-Larkin-Ovchinnikov phase can coexist in a narrow range of parameters. Motivated by these results, we establish whether this is the case also in the considered systems, \textit{i.e.}, strictly one-dimensional with the incompatibility of Fermi surfaces provided by the mass imbalance. For this, we search for non-trivial correlations between momenta of opposite-spin particles in a range of moderate mass ratios $\mu_1<\mu<\mu_2$ for which $\lambda_0$ and $\lambda_1$ are comparable. This is demonstrated in Fig.~\ref{fig4}. When the probability density $\gamma_0\left(Q\right)$ calculated for one orbital is maximal near $Q=0$, the probability density $\gamma_1\left(Q\right)$ calculated for another orbital comprises two maxima located on opposite sides and at the same non-zero distance from $Q=0$. It means that for the moderate mass ratios, the two-component fermion mixture is in a fragmented state, in which a two-body reduced density matrix $\rho_2$ can be approximated as a linear combination of two projection operators \cite{2016SakmannNatPhys}. In other words, different pairing mechanisms are simultaneously active, and different condensates of Cooper-like pairs coexist in the system. It should be emphasized that these momentum correlations cannot be simultaneously found for minor as well as major mass ratios, even if two-body orbitals $\ket{\lambda}$ with negligible probabilities $\lambda$ are taken into account.

\section{Concluding remarks}
We investigated the evolution of the ground-state properties of a two-component fermion system with a mass ratio of components $\mu$. In agreement with previous studies \cite{2019PecakPRA}, we observed that one eigenstate of a two-body reduced density matrix $\rho_2$ is dominant for a minor mass imbalance $\mu$. Therefore, the matrix $\rho_2$ can be approximated as a single projection operator, while the mixture can be described as a condensate of Cooper-like pairs with a negligible center-of-mass momentum. We also showed that the dominance of other two-particle orbital is established for a major mass imbalance. Studies of correlations between momenta of atoms revealed that in this case, the pairing mechanism is not conventional. More specifically, the Fulde-Ferrell-Larkin-Ovchinnikov phase -- in which Cooper-like pairs move with a non-zero momentum -- is supported. This momentum is a linear function of a mass ratio. Additionally, it is independent of the strength of attractive interactions. Finally, we demonstrated that conventional and unconventional pairing mechanisms can coexist for the intermediate mass imbalances. In this case, the mixture is in a fragmented state, in which a two-body reduced density matrix can be approximated as a linear combination of two projection operators.

\acknowledgements
The authors would like to thank Konrad Kapcia, Maciej Lewenstein, Piotr Magierski, and Daniel P{\c e}cak for their fruitful comments and inspiring questions at different stages of this project. 
This work was supported by the (Polish) National Science Center Grant No. 2016/22/E/ST2/00555 (TS).

\bibliographystyle{apsrev4-1}
\bibliography{_Biblio}

\end{document}